\documentclass[twocolumn,showpacs,preprintnumbers,amsmath,amssymb]{revtex4}
\usepackage{graphicx}
\input{epsf}

\newcommand{\sig}{\mbox{\boldmath{$\sigma$}}}

\usepackage{graphicx}

\input{epsf}

\newcommand{\Sig}{ \mbox{\boldmath{$\Sigma$}}}
\newcommand{\lsim}{\widetilde{<}}

\begin{document}

\date{\today}

\title{Backscattering in a 2D topological insulator and conductivity of a 2D strip.}



\author{M.V. Entin$^{1,2}$~\thanks{e-mail: entin@isp.nsc.ru}, L.I. Magarill$^{1,2}$}
\affiliation{${}^{1}$Institute of Semiconductor Physics, Siberian Branch of the Russian Academy of Sciences, Novosibirsk, 630090, Russia\\ ${}^{2}$ Novosibirsk State University, Novosibirsk, 630090, Russia}\date{\today}

\begin{abstract}A strip of 2D  HgTe  topological insulator  is studied. The same-spin edge states in ideal system propagate in opposite directions on different sides of the strip and do not mix by tunneling. Impurities, edge irregularities, and phonons produce transitions between the contra-propagating edge states on different edges.  This backscattering determines the conductivity  of an infinitely long strip. It is found that the conductivity exponentially grows with the strip width. The conductivity at finite temperature is determined within the framework of the kinetic equation. In the same approximation the non-local resistance coefficients  of 4-terminal strip are found. At low temperature the localization occurs and 2-terminal conductance of long wire vanishes, but with the exponentially long (with respect to the strip width) localization length. The transition temperature  between kinetic and localization behaviors has been found. \end{abstract}

\maketitle

\section*{Introduction}
Topological insulator (TI) is a novel actively developing  field of the solid state physics (see, {\it e.g.}, reviews \cite{has,qi} and references therein).
The main property of TI  is topological protection of the edge states that is the spin conservation together with the direction of propagation. As a consequence, the nonlocal transport appears and the conductance at zero temperature is quantized.

The topological protection (TP) is rigorous consequence of the time reversibility. In mathematical formulation single-electron elastic backscattering processes are forbidden due to conservation of $Z_2$ topological index in the systems with odd number of edges \cite{kane1,kane2,xu}, in particular in the case of a single edge. The single-edge states stay robust against not only elastic scattering but against the inelastic phonon scattering \cite{budich} both for non-interacting electrons and for Tomonaga-Luttiger liquid. On the contrary, the inclusion of the random Rashba spin-orbit coupling together with the e-e interaction opens the backscattering channel in intraedge e-e scattering \cite{strom2}. The intraedge e-e backscattering also appears due to k-dependent Rashba interaction \cite{glazm3}. Another variant of non-magnetic intraedge backscattering  is two-particle impurity scattering \cite{lezmy}.
In macroscopic 2D TI the  backscattering appears  with participation of electron puddles inside the sample \cite{glazm1,glazm2}. All these inelastic processes manifest themselves at finite temperature. However,  the elastic transitions between contra-propagating states can occur if the system possesses multiple edges (at least two on the different sides of the strip), e.g. in the region of the  strip constriction owing to non-adiabatic tunneling \cite{strom}.

In wide enough strips the elastic interedge backscattering processes are weak. However, they exist due to disorder. Till now there are no papers considering the disorder-induced interedge transitions.

The experimental evidence of the edge (and quantized) character of the transport in macroscopic HgTe quantum wells was presented in \cite{mollen}. The destruction of the quantized conductance by a weak magnetic field shows that  these properties are clearly connected with the time reversibility. The experiments on the local and non-local conductance of 2D HgTe TI have been done recently by \cite{kvon1,kvon2}. The authors demonstrate that the backscattering length in TI achieves  macroscopic values up to 1 mm. At the same time authors consider that the violation of the topological protection can be caused by the spin-flip processes.

In the present paper we study the  free-electron inter-edge backscattering in a narrow 2D TI strip caused by non-magnetic impurities, edge imperfectness and phonons and the influence of the backscattering on the conductivity.
The paper is organized as follows. First, we find the electron states in a clean strip. Then, we  consider the scattering of the edge electrons. We will use two approaches to the problem. Firstly, we will study the problem in the framework of the kinetic equation. The contra-propagating edge  states are assumed as basic states for the kinetic equation. The backscattering mean free time determines the conductivity of the infinitely long system. We examined the backscattering mechanisms caused by the impurities, the border imperfectness and phonons. Then we studied the non-local 4-terminal resistance of the TI strip. The kinetic equation approach is valid if the phase coherence between collisions are destroyed. The limitation of the validity of the kinetic approach due to dephasing intraedge forward scattering is found.  At low temperature the other approach  based directly on the single-electron Schr\"odinger equation in  a disordered system is needed.  Basing on this approach we numerically calculate the conductance using the probability of transmission through the finite-length strip. And after that we  discuss the results.

 We will neglect the e-e interaction that is justifiable if the e-e interaction constant is small.

\section*{Problem formulation}

The considered system is a strip of 2D TI from CdTe/HgTe/CdTe quantum well (see Fig.1).
 \begin{figure}[h]\label{fig1}
\centerline{\epsfxsize=7cm\epsfbox{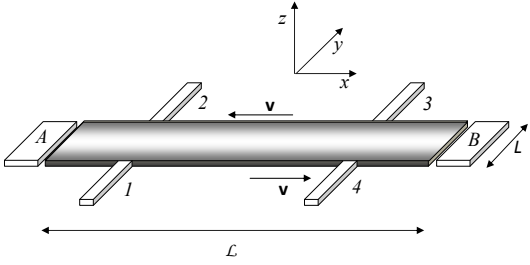}}
\caption{Sketch of the TI strip. Edge states are darkened. The directions of the velocity ${\bf v}$ correspond to $\Sigma_z=1/2$. (For $\Sigma_z=-1/2$ the directions are opposite). A and B are contacts for 2-terminal conductance measurement, 1-4 are contacts for 4-terminal measurement of non-local conductance.}
\end{figure}
 The strip in $(x,y)$ plane is determined by inequalities $-{\cal L}/2<x<{\cal L}/2$,  $-L/2<y<L/2$, ${\cal L}\gg L$. We suppose zero boundary conditions on the edges $y=\pm L/2$  and periodic conditions on $x=\pm{\cal L}/2$.
Our study is based on the effective $2\times 2$  edge-states Hamiltonian
\begin{equation}
\hat H_0 = v \hat{\sigma}_zk_x,
\label{hamiltonian107}
\end{equation}
 where   $\sig$ is the Pauli matrix, $\hbar=1$.
The Hamiltonian Eq.(\ref{hamiltonian107}) can be deduced \cite{zhou}  from the initial 2D Hamiltonian  for a CdTe/HgTe/CdTe quantum well \cite{bern}:
  \begin{eqnarray} {\cal H}({\bf k})=\left( \begin{array}{cc} H({\bf k}) & 0 \\ 0 & H^*({-\bf k})\end{array}\right),\nonumber \\ \label{Hamilt}~~H({\bf k})=\epsilon_k + {\bf d}\sig, \end{eqnarray}
   where $\epsilon_k=-D(k_x^2+k_y^2),~ d_x=Ak_x,~ d_y=Ak_y,~ d_z={\cal M}(k) = M-B(k_x^2+k_y^2). $ Parameters $A, B, D, M$ are determined by the material parameters and the thickness of the quantum well.   The upper and lower blocks of the Hamiltonian belong to the Kramers-degenerate states $j_z=1/2, 3/2$ and $j_z=-1/2, -3/2$ of the  4-fold state $j=3/2$ of the bulk HgTe zero-momentum point. These blocks can be numerated by the spin quantum number $\Sigma_z=\pm 1/2$ corresponding to the spin $\Sig$ degree of freedom. Owing to the  Kramers degeneracy it is sufficient  to solve the Schr\"{o}dinger equation for the upper block of Eq.(\ref{Hamilt}) corresponding to $\Sigma_z=1/2$. In the case of small longitudinal  momenta $k_x$ (the axis $x$ is chosen along the strip)  and large enough width $L$ one can write for the  energy spectrum and wave functions \cite{zhou}:
\begin{eqnarray}
E_\sigma(k_x) = \sigma vk_x  \label{E};     ~~~~~~(\sigma=\pm 1)\\
   \Psi_{k_x;\sigma}(x,y) = \frac{e^{ik_xx}}{\sqrt{{\cal L}}}\psi_{\sigma}(y),\nonumber\\ \psi_{\sigma}(y)=\tilde{c}_{\sigma}g_{\sigma}(y)(1,-\sigma\eta), \label{Psi}\\
    \nonumber g_{\sigma}(y)=(f_+(y)- \sigma f_-(y)).
\end{eqnarray}
Here energies $E_{\sigma} $ are counted from $E_0=-MD/B$,  $v=A\sqrt{B_+B_-/B^2},   ~ B_\pm=B\pm D$, $\tilde{c}_{\sigma}$ are  the normalization constants. Expressions for $f_\pm(y)$ are given by Eqs.(7,8)  in \cite{zhou}.
Eqs.(\ref{E}),(\ref{Psi}) were found  by solving the Shr\"{o}dinger equation with zero boundary conditions at $y=\pm L/2$:
\begin{eqnarray}
  [M-B_+(k_x^2-\partial_y^2)]\psi_1+A(k_x-\partial_y)\psi_2 &=& E\psi_1 \\
    A(k_x+\partial_y)\psi_1-[M-B_-(k_x^2-\partial_y^2)]\psi_2 &=& E\psi_2,
\end{eqnarray}
where $\psi_{1,2}(y)$ are  the components of the two-component spinor $\psi(x,y)=(\psi_1(y),\psi_2(y)).$

In the limit of large $L$ and small $k_x$ expressions for functions $ g_{\sigma}(y)$  are simplified and given by:
\begin{eqnarray} \label{wf5}                    
g_{\sigma}(y) \simeq 2[e^{-\lambda_1 L/2-\sigma \lambda_1y}-e^{-\lambda_2 L/2-\sigma \lambda_2y}].
    \end{eqnarray}

    The number $\sigma$ is conserved in a clean system and one can consider $\sigma/2$  as a pseudospin. The  functions $ \psi_{\sigma}(y)$  exponentially decay from the edges $y=-\sigma L/2$, correspondingly.  Functions with different $\sigma$ are weakly overlapped with each other if $\lambda_{1,2}L\gg 1$. In fact, the wave functions (\ref{wf5}) and linear spectrum (\ref{E}) correspond to  insulated edges. This approximation is valid for large enough electron energy exceeding  the gap $\Delta$ \cite{zhou}: $$|E|\gg\Delta = 4\frac{|AB_+B_-M|}{\sqrt{B^3(A^2B-4B_+B_-M)}}e^{-\lambda_2L}.$$ Due to the exponential  decay  of $\Delta$ with increase of $L$ this limitation can be easily fulfilled.

  The presence of disorder (impurities, edge roughness, phonons)  leads to  transitions between edge   states with different $\sigma$. The decay rate $\lambda_1$ is larger than $\lambda_2$. In the limit of large $L$ this permits to keep only one exponent with $\lambda_2$ in $g_{\sigma}(y)$ when one calculates  overlapping integrals.

In the same approximation we have for $\eta, \ \tilde{c}_{\sigma}$, and $\lambda_{1,2}$
\begin{eqnarray} \eta^2=\frac{B_+}{B_-}, ~~
   \tilde{c}_{+1}^2\simeq \tilde{c}_{-1}^2\equiv \tilde{c}^2= \frac{AMB_-\sqrt{B_+B_-}}{4B (A^2B-4MB_+B_-)}, \\
  \lambda_{1,2} = \frac{A}{2\sqrt{B_+B_-}}\pm \sqrt{\frac{A^2}{4B_+B_-}-\frac{M}{B}}\\
   (\lambda_{1}+\lambda_{2}= \frac{A}{\sqrt{B_+B_-}}, ~~~~ \lambda_{1}\lambda_{2}=\frac{M}{B}).\nonumber
\end{eqnarray}

  The Hamiltonian (\ref{hamiltonian107}) results from the 2D Hamiltonian (\ref{Hamilt}) when $\lambda_{1,2}\gg k_x,1/L$. It should be complemented by the potential of interaction with defects. In the same representation the potential is given by the $2\times 2$ matrix $\hat{U}(x)$ with matrix elements being equal to $U_{\sigma'\sigma}(x)=\langle\psi_{\sigma'}(y)|U(x,y)|\psi_{\sigma}(y)\rangle$ which are composed by a projection of the potential to the states $\psi_{\sigma}(y)$.  This matrix depends on the coordinate $x$ only.
  The total edge-state Hamiltonian reads
 \begin{equation}
\hat{H} = -iv \hat{\sigma}_z\partial_x+\hat{U}(x).
\label{hamiltonian}
\end{equation}

\section*{Impurity scattering}
In the kinetic approach the conductivity is caused by the transitions  of electrons between the edge states. In this section we will find the transition probability under scattering on impurities located inside the strip.
The potential energy of interaction of an electron with impurities is given by

\begin{equation}\label{U}
    U_{imp}=\sum_n u_n({\bf r}) =\sum_n u({\bf r}-{\bf r}_n)=\sum_{{\bf q},n} \tilde{u}_{\bf q}e^{i{\bf q}({\bf r}-{\bf r}_n)},
\end{equation}
where  $\tilde{u}_{\bf q}=\int u({\bf r})e^{-i{\bf qr}}d{\bf r}/S$ is the Fourier transform of the potential of an individual center, $S=L{\cal L}$ is the area of the system.

We will be interested in transitions with backscattering.  Necessary  matrix elements of corresponding matrix $\hat{U}(x)$ can be written as
\begin{eqnarray}\label{U matr el}
    U_{+1;-1}(x)=U_{-1;+1}(x)=  \tilde{c}^2(1-\eta^2)\times \nonumber \\ \sum_{{\bf q},n}\tilde{u}_{\bf q}e^{iq_x(x-x_n)-iq_yy_n}\int_{-L/2}^{L/2} dy g_{+1}(y)g_{-1}(y) e^{i{q_yy}}.
    \end{eqnarray}
 Again in the case of large $L$ one can find approximate expression for $\int dy g_{+1}(y)g_{-1}(y)e^{iq_yy} $
   at   $\lambda_1>\lambda_2$:
\begin{equation}\label{inty}
   \int dy g_{+1}(y)g_{-1}(y)e^{iq_yy}\simeq \delta_{q_y,0}4 Le^{-\lambda_2L}.
    \end{equation}
As a result for interaction of an electron with an individual impurity we have
\begin{eqnarray}\label{u matr el}
    u_{\sigma;-\sigma}(x)= 4Le^{-\lambda_2L}\tilde{c}^2(1-\eta^2) \sum_{q_x}\tilde{u}_{q_x,0}e^{iq_xx}.
    \end{eqnarray}

    Using Eq.(\ref{u matr el}) we can write  the transition probability between states  $|k_x';\sigma\rangle$ and $|k_x';-\sigma\rangle$ as follows:
     \begin{eqnarray}\label{Wimp}
        W_{k_x',-\sigma;k_x,\sigma}^{(imp)}=\nonumber \\ 32\pi N\tilde{c}^4(1-\eta^2)^2|\tilde{u}_{-2k_x,0}|^2L^2e^{-2\lambda_2L}\delta(v(k_x'+k_x)),
 \end{eqnarray}
where $N$ is the total number of scattering centers. In Eq.(\ref{Wimp}) average over distribution of impurities have been  carried out. Eq.(\ref{Wimp}) can be presented in the form:
\begin{equation}\label{Wimp1}
    W_{k_x',-\sigma;k_x,\sigma}^{(imp)}=\frac{\pi }{{\cal L}\tau}\delta(k_x'+k_x),
\end{equation}
where we have introduced the  relaxation time due to impurity scattering:
\begin{eqnarray}\label{tauimp}
\frac{1}{\tau} =\frac{8n_s}{v}|\bar{\tilde{u}}_{2k_x,0}|^2Le^{-2\lambda_2L} \lambda_2^2~a, \\ a=\left[{\frac{1-\eta^2}{1+\eta^2}}
  \frac{\lambda_1(\lambda_1+\lambda_2)}{(\lambda_1-\lambda_2)^2}\right]^2.\nonumber
\end{eqnarray}
Here $\tilde{u}_{\bf q}=\bar{\tilde{u}}_{\bf q}/(L{\cal L}), ~~ n_s=N/S$ is the impurity concentration.

\section*{Scattering on edge imperfections}
Let the edges to be imperfect, namely having  shapes $y=\sigma L/2+h_{\sigma}(x)$, where $h_{\sigma}(x)$ are random functions with correlators  $\langle h_{\sigma}(x) h_{\sigma'}(x') \rangle=w_\sigma\delta(x-x') \delta_{\sigma,\sigma'}$.
Interaction of an electron with roughness of edges is determined by the pseudo-potential \cite{prang}, \cite{chaplik}:
\begin{equation}\label{Uedge}
    U_{edge} = \frac{1}{2m}\sum_\sigma h_{\sigma}(x)\hat{k}_y
\delta(y+\sigma L/2)\hat{k}_y.
\end{equation}
In analogy with the impurity case we obtain for the transition probability caused by the edge imperfectness Eq.(\ref{Wimp1}) with replacement of the relaxation time by
\begin{eqnarray}
\frac{1}{\tau} =\frac{8}{m^2v}(w_{+1}+w_{-1})e^{-2\lambda_2L}\lambda_2^4(\lambda_1-\lambda_2)^2a.
\end{eqnarray}
The order of $w_\sigma$ is determined by the typical height $h_0$ and width $w_0$  of roughness: $w_\sigma\sim h_0^2w_0$.

\section*{Conductivity}
Let us consider a long strip in a longitudinal  external electric field $\cal E$. The linearized kinetic  equation for the electrons of edge states is:
\begin{equation}\label{Kineq}
   \sigma e{\cal E}v\frac{\partial f_0(E_\sigma(k_x))}{\partial E}=\sum_{k_x'}W_{k_x',-\sigma;k_x,\sigma}(\chi_{\sigma,k_x'}-\chi_{\sigma,k_x}).
\end{equation}
Here $f_{\sigma,k_x}0(E_\sigma(k_x))+\chi_{\sigma,k_x}$ is the distribution function, ~~~~$f_0(E_\sigma(k_x))$ is the Fermi function.  Eq. (\ref{Kineq}) is easily solved using  the identity $\chi_{-\sigma,-k_x}=-\chi_{\sigma,k_x}$:
     \begin{equation}\label{chi}
    \chi_{\sigma,k_x}=-\sigma e{\cal E}\frac{\partial f_0(E_\sigma(k_x))}{\partial E}v\tau,
           \end{equation}
    where   $\tau$  is the relaxation time. For    $1/\tau$ one should utilize  the sum of scattering rates due all to considered  mechanisms.
As a result, we obtain the classical  conductivity of the  degenerate electron gas $G_0l$ and corresponding conductance of a finite strip $G_0l/{\cal L}$ expressed via the conductance quantum $G_0=2e^2/h$  and the mean free path at the Fermi energy $l=v\tau$.
\section*{Localization at low temperature}

Above we have considered the backscattering problem within the framework of the kinetic equation. This approach is valid in the case of destroyed phase coherence. In the low-temperature limit the quantum approach for electron propagation is needed. Here we consider the problem   basing on the localization theory \cite{mott}.

In absence of the potential the solutions of the stationary Schr\"odinger equation $(\hat{H}-E)\xi=0$ are $\xi_{k,\sigma}=e^{ikx}(1+\sigma,1-\sigma)/2$  with the energy $E_{k,\sigma}=\sigma vk$.

The matrix of potential $\hat{U}(x)$ mixes the states of different edges.
With account for the diagonal elements of the potential  $U_{\sigma,\sigma}(x)$ only, the stationary solutions with the same $E_{k,\sigma}$ convert to  \begin{eqnarray}\xi_{k,\sigma}=e^{ikx}\frac{1}{2}\left(\begin{array}{c}
                  1+\sigma\\
                 1-\sigma \\
                \end{array}\right)\exp\left(-i\frac{\sigma}{v}\int U_{\sigma,\sigma}dx\right)\label{wf}.\end{eqnarray}
The wavefunctions (\ref{wf}) contain  the phase corrections and do not contain the backscattering.  To account for the backscattering one should include small non-diagonal elements  $U_{\sigma,-\sigma}$ ($U_{\sigma,-\sigma}\ll U_{\sigma,\sigma}$)  into the consideration.

Let us consider a single impurity located in the point $(x_n, y_n)$ with the potential $\hat{u}(x-x_n)$.

Off-diagonal matrix elements of $\hat{u}(x)$ follow from  Eq.(\ref{U matr el}), while diagonal elements give inessential corrections to the phases.

An electron incoming from $-\infty$ in the edge $\sigma=1$ collides with the impurity and can pass to the edge $\sigma=-1$ and reverse the momentum $k\to -k$ and the direction of motion. Alternatively, the electron can pass to $\infty$ conserving $k$ and $\sigma$.

To find the amplitude of these processes  one should solve the stationary Schr\"odinger equation  with the boundary conditions $\xi=(e^{ikx},re^{-ikx})$ at $x\to-\infty$ and $\xi=(te^{ikx},0)$ at $x\to\infty$, where $t$ and $r$ are amplitudes of transmission and reflection, accordingly. The unitarity yields $|r|^2+|t|^2=1$.

 In particular, solving the Schr\"odinger equation in the first Born order we have

\begin{equation}\label{U2}
    r=(1/v)\int_{-\infty}^\infty e^{2ikx}u_{+1,-1}(x)dx,
\end{equation}
In this approximation $r\ll 1,~t\approx1$.

Now go to the problem of many impurities situated in the points $(x_n,y_n)$. Here we will not restrict ourselves by the Born case and consider the reflection amplitudes $r_n$ as given arbitrary real numbers between 0 and 1. The wave function between n-th and (n+1)-th impurities  ($x_n<x<x_{n+1}$) is
\begin{equation}
(a_ne^{ik(x-x_n)},
b_ne^{-ik(x-x_n)}).
\end{equation}
We assume here that between impurities  electrons  freely propagate. This requires the absence of impurity potential  overlapping: the mean distance between impurities  along $x$-axis $(n_sL)^{-1}$ is larger than the characteristic size of impurity potential.
The model under consideration is illustrated by the Fig.2
\begin{figure}[h]\label{fig2}
\centerline{\epsfxsize=7cm\epsfbox{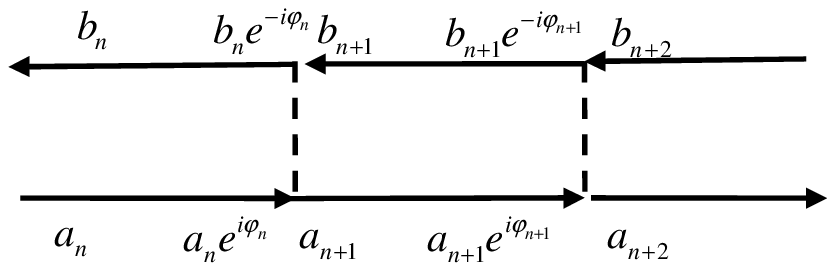}}
 \caption{One-dimensional model of electron localization on the edge states.}
\end{figure}
 In the points $x_n$ electrons meet the n-th impurity and experience backscatterings.  The scattering by the impurities determines the system of algebraic equations for $a_n, ~b_n$:
\begin{eqnarray}\label{loc1}b_ne^{-i\phi_n}=r_na_ne^{i\phi_n}+\sqrt{1-|r_n|^2}b_{n+1},\nonumber\\
a_{n+1}=\sqrt{1-|r_n|^2}a_ne^{i\phi_n}+r_nb_{n+1}.\end{eqnarray}
 Here $\phi_n=k(x_{n+1}-x_n)$.\footnote{Note, that the diagonal elements of the potential $u_{\sigma,\sigma}$ also can be included  into consideration; however, they give corrections to the phases of transmission and reflection which can be accommodated into  $\phi_n$. The same is valid with respect to the potential of {\it all} impurities $U_{\sigma,\sigma}$.}.

 The positions of impurities are random, hence we can consider $\phi_n$ as randomly distributed numbers.  For modeling we should assume that $\phi_n$ are randomly distributed within the range $(0,2\pi)$. This assumption is valid at least if $x$-distance between subsequent impurities exceeds $2\pi/k$.  Below we set  $r_n\equiv r$.

 In a matrix form Eq.(\ref{loc1}) reads
 \begin{equation}
\left(
   \begin{array}{c}
     a_{n+1} \\
     b_{n+1} \\
   \end{array}
\right)=
S_n\left(
   \begin{array}{c}
    a_n \\
     b_n \\
   \end{array}
\right),
\end{equation}
 where
\begin{equation}
S_n=\frac{1}{\sqrt{1-r^2}}\left(
   \begin{array}{cc}
    e^{i\phi_n}(1-2r^2) &re^{-i\phi_n} \\
     -re^{i\phi_n} & e^{-i\phi_n} \\
   \end{array}
\right).\end{equation}
Consider a finite strip with $N=n_sL{\cal L}$ impurities in it. To find the transmission coefficient of the total system we will act like \cite{mott}. Namely, we start from $a_1=0,b_1=1$ ($a_1=0$ means no incident wave at $n=1$, $b_1=1$ means normalized to unity intensity of the wave transmitted to the left).
Then $|b_N|^{-2}$ gives the transmission coefficient, $G_0|b_N|^{-2}$ is the conductance. The resulting recurrence is
\begin{equation}
\left(
   \begin{array}{c}
     a_N \\
     b_N \\
   \end{array}
\right)=
\left(\prod_{n=1}^NS_n\right)\left(
   \begin{array}{c}
    0 \\
     1 \\
   \end{array}
\right).
\end{equation}
In a large system $\ln |b_N|\propto N$ and $$(n_sL)^{-1}\lim_{N\to\infty}N/(2\ln|b_N|)$$ determines the localization length.
Fig.3 represents the calculated inverse localization length (in unites of $(n_sL)$) versus the amplitude of reflection $r$. The dependence is approximately quadratic, in accordance with the mean free path calculated in the Born approximation.
\begin{figure}[h]\label{fig2}
\centerline{\epsfxsize=8cm\epsfbox{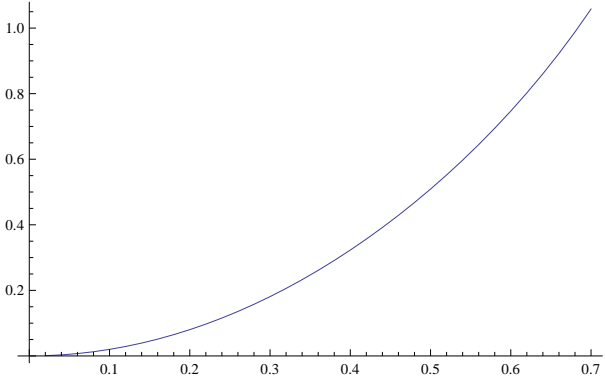}}
 \caption{Inverse localization length (in unites of $n_sL$) versus the amplitude of reflection $r$.}\end{figure}

\section*{Phonon mechanisms of electron backscattering and  dephasing}

 The impurity scattering conserves the phase coherence and hence, strictly speaking, can not be considered within the kinetic equation approach. However, this is not the case when any decoherence factor is taken into account. Strong enough decoherence revives the kinetic equation applicability.

 Unlike the impurities, the phonons do not conserve the phase coherence. Below we consider the scattering of electrons
 by phonons. Note that at large enough temperature, the phonon backscattering can be considered in the same way as  the impurity scattering by neglecting the emitted phonon frequency as compared with the temperature.
The Hamiltonian of electron-phonon interaction is
 \begin{equation}\label{Heph}
   H_{e-ph}=\sum_{k,\sigma,k'\sigma',\bf q}c_{\bf q}J_{q_z}J_{q_y;\sigma'\sigma}a_{\sigma',k'}^+a_{\sigma,k}b^+_{\bf q}\delta_{k'-k,q_x} + h.c.
\end{equation}
 Here $a_{\sigma,k}^+,~a_{\sigma,k}$ are the edge electron creation/annihilation operators,  $b_{\bf q}^+,~b_{\bf q}$ are the  creation/annihilation operators of bulk longitudinal acoustic phonons with 3D momentum ${\bf q}$. Quantities  $ c_{\bf q}, ~ J_{q_z}$ and $J_{q_y;\sigma\sigma'}$ are given by
 \begin{equation}\label{Heph2}
  c_{\bf q}=\frac{\Lambda q}{\sqrt{2\rho \omega_{\bf q}\Omega}},~~~ J_{q_y;\sigma'\sigma}=\langle \sigma'| e^{iq_y y}|\sigma\rangle,, ~~ J_{q_z}=\int dz \zeta^2(z), \end{equation}
 where $\Lambda$ is the deformation potential constant, $\rho$ is the crystal density, $\Omega$ is the system volume, $\omega_{\bf q}=cq$ is the phonon frequency ($c$ being the sound velocity), $\zeta(z)$ is the  ground state wave-function of the quantum well CdTe/HgTe/CdTe.
In our consideration it is assumed that electrons interact by the deformation potential  with the bulk longitudinal acoustic phonons only; the difference between HgTe and CdTe elastic constants and deformation potentials is neglected.

Similar to Eq.(\ref{Wimp}) we have found the interedge ($\sigma\to -\sigma$) transition probability caused by the phonons.  As a result, instead of Eq.(\ref{tauimp}) we have for the backscattering time
 \begin{equation}\label{tauph11}
   \frac{1}{\tau}=\frac{8\Lambda^2L T\lambda_2^2 e^{-2\lambda_2L}}{v\rho c^2}a\int \zeta^4(z)dz ,
\end{equation}
 The integral in Eq.(\ref{tauph11}) has the order of the inverse quantum well width $1/d$. (It should be noted  that in contrast to the impurity scattering the scattering due to phonons demanded accounting for the transversal (i.e. along $z$) structure of the edge state.) In deducing  Eq.(\ref{tauph11}) we have utilize the condition that the temperature $T$ is much larger than the  characteristic frequency of emitted phonons $c/d$.
The phonon  backscattering grows with the temperature and has the same smallness caused by the overlapping of edge states  as the impurity scattering. Note
 that the electron-phonon scattering rate Eq.(\ref{tauph11})  can be found by replacement  the total number of impurities by the phonon distribution function and corresponding replacement of the interaction constant. For conductivity one should collect the relaxation rates caused by impurities, edges and phonons together.

 Now go to the forward phonon-induced scattering $\sigma\to \sigma$. This process being essentially stronger than the backscattering conserves the electron velocity and hence does not affect  the kinetics.  The role of  the forward scattering is to control the phase coherence in the system, namely electron dephasing time $\tau_{\phi}$.
Electrons propagating along an isolated edge obtain random phases due to the phonon field. Here we will discuss the phonon mechanism in the low-temperature limit when the frequency of emitted/absorbed phonons has the order of electron thermal energy $T$ (in other words, the process is inelastic) and, hence, $\tau_\phi$ coincides with the inelastic forward relaxation time \footnote{Note, that strict consideration of the phase coherence requires the solution of the quantum kinetic equation for the density matrix in $\sigma$.  Strong dephasing, however,  suppresses the off-diagonal elements of this matrix. As a result, the diagonal elements of the density matrix  become larger than off-diagonal elements; that revives the classical kinetic equation.}.
Noting, that  the main contribution to $\tau_\phi$  arises from the  forward scattering, we find \begin{eqnarray}
\tau_\phi^{-1}=2\pi\sum_{\pm,{\bf q},k'}|c_{\bf q}|^2|J_{q_z}|^2|J_{q_y;+1,+1}|^2\delta_{k',k\mp q_x}\times \nonumber \\(N_{\bf q}+1/2\pm 1/2)(1-f_0(E_{+1}(k'))) \delta(vk-vk'\mp cq).
   \end{eqnarray}
Here $N_{\bf q}$ is the Bose-Einstein distribution function. Using the relations  $cq\sim T$, $q_x\approx cq/v\ll q$, $q_y\lsim \lambda_{1,2}$, $q_x\ll q_y< q$, we obtain at $T\to 0$ $J_{q_z}\approx 1, ~J_{q_y;\sigma'\sigma} \approx 1$. Then
        \begin{eqnarray}\label{tau-phi2}
   \tau_\phi^{-1}=\frac{\Lambda^2T^3}{4\pi\rho c^4v}\int_0^\infty x^2dx\frac{\coth{(x/2)}(e^{2\epsilon}+e^{\epsilon})}{(e^{\epsilon-x}+1)(e^{\epsilon+x}+1)},
   \end{eqnarray}
where $\epsilon=(E-E_F)/T$. The expression (\ref{tau-phi2}) is valid if $T\ll \lambda_2 c$.  Value of $\tau_\phi$ depends on the electron energy. The  integral in (\ref{tau-phi2}) runs from $\approx 4.2$ at $\epsilon=0$ to $\epsilon^3/3$ at $\epsilon\rightarrow\infty$. Value of $\tau_\phi$ averaged with the derivative of the Fermi function $-e^{\epsilon}/(e^{\epsilon}+1)^2$  gives \begin{eqnarray}\label{tau-phi24}\langle\tau_\phi\rangle=0.354\frac{4\pi\rho c^4v}{\Lambda^2T^3}.\end{eqnarray}
Eq. (\ref{tau-phi24}), in fact, gives the mean free time of inelastic forward scattering which is reasonable estimation for $\tau_\phi$. One can see that at low temperature $\tau_\phi$ grows $\propto T^{-3}$, but do not contain an exponentially large factor caused by the wave function overlapping.

The validity of the kinetic regime needs $\tau_\phi(T)\ll \tau$, {\it vice versa} the localization occurs if $\tau_\phi(T)\gg \tau$; the transition between these regimes occurs when $\tau_\phi(T)\sim\tau$, where $\tau$ in the low-temperature limit does not depend on $T$.  More detailed  consideration goes beyond the scope of this paper.

\section*{Nonlocal conductivity}

The currents in the TI are localized near the strip edges. The edge states determine one-dimensional "quantum wires", hence the conductance has non-local character. The nonlocality of the transport means that the voltage applied to a close pair of contacts (for example, contacts 1 and 2 in the Fig. 1) penetrates to the distance much larger than the distance between them. Let us consider the non-local conductance at finite temperatures within the framework of the kinetic equation. Instead of solving the problem of conductivity we will deal with the diffusion keeping in mind the Einstein relation between the conductivity and diffusion coefficient. Let an infinite strip has 4 contacts (see Fig.1) located at points $(X_i,Y_i)$, $X_1=X_2=0$, $Y_1=Y_4=-L/2$, $Y_2=Y_3=L/2$, $X_3=X_4$. We assume that  the contacts 1 and 2 are current contacts and contacts 3 and 4  are potentiometric.

The application of the voltage between contacts 1 and 2 can be interpreted as difference between chemical potentials of emitted (absorbed) electrons $E_F-eV_{21}/2$ and $E_F+eV_{21}/2$. The equations for the distribution functions read   $$\sigma v\partial \chi_{k,\sigma}(x)/\partial  x=(\chi_{-k,-\sigma}(x)-\chi_{k,\sigma}(x))/2\tau.$$
The boundary conditions are $\chi_{k,\sigma}(\pm\infty)=0$,  $\chi_{k,\pm 1}(x=X_1 \pm 0)=\mp(eV_{21}/2)\partial f_0(vk)/\partial E_F$.

Solving the equations we find the current $J_{21}=G_0V_{21}$ and the potential between the edges in the distance $x$ from the current contacts: $V(x)=V_{21}e^{-|x|/l}$. The difference of the potentials between contacts 3 and 4 is $V_{34}=V_{21}e^{-|X_{32}|/l}$, where $X_{ij}=X_i-X_j$. Denote by $R_{i,j;k,l}=V_{k,l}/J_{i,j}$ the resistance  between k and l contacts if the current $J_{i,j}$ is applied between contacts  i and j. We find that the resistance $R_{1,2;1,2}=G_0$, $R_{3,4;1,2}=e^{-|X_{32}|/l}/G_0$. Hence, the characteristic damping length of non-locality is the mean free path. Exponentially large value of $l$ determines essential non-locality of the conductivity.

\subsection*{Numerical estimations}
Let us present  numerical estimation of the free path for the case of  scattering by the bare Coulomb impurities. In this case $\bar{\tilde{u}}_{\bf q}=2\pi e^2/(\kappa q)$, where $\kappa$ is the dielectric constant of the surrounding medium (CdTe in our case).  For $l$ we have
$$l=\frac{1}{8\pi^2\lambda_2}\frac{E_F^2\kappa^2}{e^4n_s}\frac{1}{a}\frac{e^{2\lambda_2L}}{\lambda_2L}.$$

For material parameters we take magnitudes from \cite{konig}:  $A=-364.5$meV nm, $B=-686$ meV nm$^2$, $D=-512$ meV nm$^2$, $M=-10$ meV,  $\kappa=10.2$. The calculated parameters are  $\lambda_1=0.78$nm${}^{-1}$, $\lambda_2=0.018$nm${}^{-1}$, $v=3.88\cdot 10^7$cm/s, $a=0.639$.
At $E_F=3$meV, $n_s=10^{11}$ cm${}^{-2}$, $L=200$ nm, we get $l=1.8$ $\mu$m. Note, that exponential dependence of the mean free path on the width leads to quick growth of this parameter with $L$.

\subsection*{Discussion and conclusions}

We have based on the topological protection of the edge states of ideal TI strip. The topological protection results in the equality of the strip conductance to the quantum independently on the strip length.  However, we demonstrated that intervention of the interedge scattering makes long TI strip to be similar to a 1D wire of usual conductor: at large temperature the strip possesses the finite conductivity and at low temperature the localization occurs. The difference with a quantum wire consists in exponentially long (with respect to the strip width) mean free path and the localization length. This understanding resolves the imaginary contradiction between the theory of localization and the topological protection. It should be emphasized that our consideration is limited by the case when the electron energy exceeds the gap caused by interaction between edges. This limitation is not critical due to exponentially small value of the gap.

Let us discuss the correspondence with the experimental measurements  \cite{kvon1,kvon2}. In experiments \cite{kvon1,kvon2} it was found that  the conductivity of a macroscopic 2D TI is: i) non-local, ii) non-quantized and iii) temperature-independent at low temperature. The case i) means the edge character of conductivity while ii) means deflection from the ballistic transport due to the backscattering. The case iii) indicates on  the impurity mechanism of the backscattering.  Very large macroscopic backscattering length in \cite{kvon1} tells  about weak influence of spin-flip scattering.  The macroscopic system in \cite{kvon1,kvon2}  is not narrow strip considered here.  In a macroscopically wide device the transitions between the strip edge are forbidden. However, we think that the real TI can have many puddles of normal semiconductor phase due to the fluctuations of the HgTe quantum well thickness. These puddles provide existence of multiple  inner edges between TI and normal  semiconductor. The chain of  transitions between these edge states shall produce transition to the other external edge with the opposite direction of motion. If the impurities (edge imperfections)  give the main contribution to this process in conditions of destroyed coherence  one may expect no temperature dependence of   the conductivity in accordance with \cite{kvon1,kvon2}. This possibility is alternative to the process discussed in \cite{glazm1,glazm2}.

In conclusions, we have solved problems of the conductivity of the infinitely long strip of the TI. At finite temperature the kinetic equation approach is valid. In this approximation the conductivity is determined by the mean free path due to the interedge backscattering on impurities, edge imperfections and phonons. The validity limit of this approach due to the dephasing caused by the phonon-induced intraedge forward scattering has been established. The non-local 4-terminal resistance coefficients have been also found. The low-temperature limit of the finite-wire conductance was examined on the basis of the localization theory.

It should be emphasized that we have neglected the spin non-conservation caused by the spin-orbit interaction, magnetic impurities and superfine interaction. These mechanisms go beyond the paper scope, but look more weak than considered ones.
\subsection*{Acknowledgements}
The work was supported by the RFBR grants 13-0212148 and 14-20-00593.

\end{document}